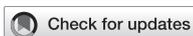

# Enhanced weak force sensing based on atom-based coherent quantum noise cancellation in a hybrid cavity optomechanical system

S. K. Singh[1†], M. Mazaheri[2†], Jia-Xin Peng[3]*, A. Sohail[4], Mohammad Khalid[1] and M. Asjad[5]

[1]Graphene and Advanced 2D Materials Research Group (GAMRG), School of Engineering and Technology, Sunway University, Petaling Jaya, Selangor, Malaysia, [2]Department of Basic Science, Hamedan University of Technology, Hamedan, Iran, [3]State Key Laboratory of Precision Spectroscopy, Quantum Institute for Light and Atoms, Department of Physics, East China Normal University, Shanghai, China, [4]Department of Physics, Government College University, Faisalabad, Pakistan, [5]Advanced Communication Engineering (ACE) Centre of Excellence, Universiti Malaysia Perlis, Kangar, Perlis, Malaysia

The weak force sensing based on a coherent quantum noise cancellation (CQNC) scheme is presented in a hybrid cavity optomechanical system containing a trapped ensemble of ultracold atoms and an optical parametric amplifier (OPA). In the proposed system, the back-action noise can be completely eliminated at all frequencies and through the proper choice of the OPA parameters, and the noise spectral density can also be reduced at lower frequencies. This leads to a significant enhancement in the sensitivity of the cavity optomechanical weak force sensor, and the noise spectral density also surpasses the standard quantum limit (SQL) even for the small input power at the lower detection frequency. Furthermore, the experimental feasibility of this scheme is also briefly discussed. This study can be used for the realization of a force sensor based on hybrid cavity optomechanical systems and for the coherent quantum control in macroscopic systems.

KEYWORDS

quantum noise cancellation, optical parametric amplifier, standard quantum limit, optomechanical force sensor, ultracold atoms

## 1 Introduction

Cavity optomechanics is an emerging research area which explores the coherent coupling between the optical mode and the mechanical mode through the radiation pressure of photons trapped inside an optical cavity [1–3]. It has also made significant advances in this modern era of quantum technology such as ultrahigh-precision measurement [4], gravitation-wave detection [5], quantum information processing (QIP) [6], non-classical photon statistics [7–10], quantum entanglement [11–19], macroscopic quantum coherence [20–22], ground-state cooling of mechanical oscillator [23, 24], and optomechanically induced transparency (OMIT) [25–32] including quantum teleportation [33, 34] and quantum communication [33–35]. Moreover, the measurement of weak forces at the quantum limit of sensitivity is of particular importance [36, 37] and also leads to major developments in cavity optomechanical sensors [38, 39]. In particular, one of the greatest applications of optomechanical sensors is the detection of gravitational waves [40], which directly leads to the rapid development of sensors based on cavity optomechanical systems [1, 41, 42]. In addition, optomechanical sensors can also give accurate results to measure various physical quantities like mass [43–45], acceleration [46–49], displacement [50–53] and force [54–58], magnetometry [59–61], and acoustic sensing [62, 63]. However, such kind of optomechanical sensors have limitations due to the presence of the shot noise and the back-action noise, and the competition between them leads to the notion of the standard quantum limit (SQL) [56, 64]. The quantum non-demolition measurement [65] through quantum entanglement [66] or squeezing [67–69] can be used to surpass the SQL, e.g., some pioneering theoretical works suggest that the squeezed states can be used to detect the gravitational waves using supersensitive interferometry [38, 70, 71]. Furthermore, a few experimental protocols to investigate possible quantum gravity effects on macroscopic mechanical oscillators that are preliminarily prepared in a high purity state are given [72].

In the optomechanical sensor, as we increase the input power to enhance the measurement strength and to decrease the shot noise, it also increases the unwanted measurement back-action noise. This means that both these noises have opposite scaling with the input laser power and we need to reduce or completely eliminate the back-action noise to increase the force sensitivity, that is why the complete elimination of back-action noise is the most important goal for all the optomechanical sensing based platforms. Except for the quantum thermometry using sideband asymmetry of a laser-cooled mechanical resonator, where back-action is not unwanted, but desirable [2, 73]. It has been shown that back-action-evading measurements of a single quadrature of nanomechanical motion can increase the force sensitivity [74], whereas two-mode back-action-evading measurements in a single-mode cavity can also surpass the SQL [75]. Another approach for obtaining the sub-SQL is based on the CQNC of the back-action noise through quantum interference. In this approach, an anti-noise path in the system is introduced through the addition of an ancillary oscillator which manifests an equal and opposite response to the light field, and it cancels the back-action noise [76–78]. So, the CQNC scheme can delete the back-action noise induced due to the radiation pressure at all frequencies and surpass the SQL [79, 80]. In recent developments, hybrid optomechanical systems containing atomic ensembles significantly improve the optomechanical cooling [81–85], the realization of quantum squeezing of the motion of mechanical oscillators [86], and provide entangled Einstein–Podolsky–Rosen (EPR) states and squeezed states [76, 87–89]. Interestingly, it is possible to enhance the precision of a position measurement inside an optomechanical cavity with the optical parametric amplifier (OPA) [90] and force sensitivity in a simple [91] and hybrid optomechanical system with trapped atomic ensemble [92, 93]. Very recently, a study on Zeptonewton force sensing with squeezed quadratic optomechanics was proposed, and the results showed that by optimizing the system, a force sensitivity seven orders of magnitude higher than any conventional linear cavity optomechanical sensors could be achieved [94].

Based on these works, the present work aims to study the CQNC scheme-based weak force sensing in a hybrid optomechanical system containing both the degenerate OPA and the trapped atomic ensemble inside the cavity. In this way, we have investigated the effects of the amplitude and phase of the OPA on the spectral force sensing. The coupling between the optomechanical cavity and the atomic ensemble is necessary for the CQNC process. On the other hand, tuning the OPA parameters can improve the precision of force sensing and reach the sub-SQL sensitivity. This system can also create wider spacing in the normal mode splitting (NMS) and a greater degree of the squeezing spectrum as compared with just an OPA or an atomic ensemble [95].

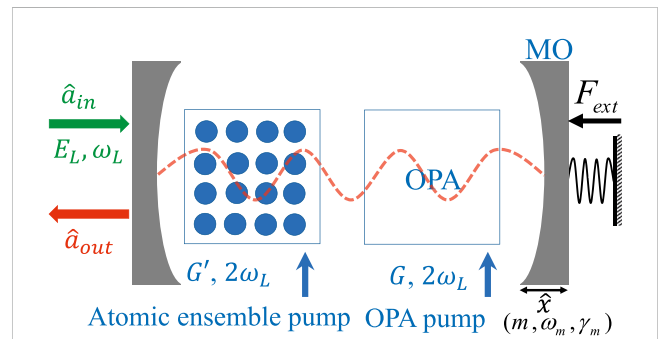

FIGURE 1
Overview of the hybrid optomechanical system which consists of an optomechanical cavity, a degenerate optical parametric amplifier (OPA), and an ensemble of two-level ultracold atoms. The atomic ensemble behaves effectively as a negative-mass oscillator (NMO) under the bosonization process [92, 93, 96]. An external force $F_{ext}$ is applied on the MO acting as a force sensor. The cavity is also driven by an external classical field with power $P_L$ and frequency $\omega_L$.

The present paper is organized as follows. In Section 2, we present the hybrid optomechanical system and its model Hamiltonian for the optomechanical force sensor. We obtain the linearized equations of motion in Section 3. In Section 4, we have discussed the enhancement of CQNC and subsequently weak force sensing under different conditions. We have also briefly discussed the experimental feasibility of this proposed scheme. Finally, we have given a conclusion in Section 5.

## 2 The model Hamiltonian

We consider the hybrid optomechanical system as shown in Figure 1, which consists of a single-mode optical cavity with resonance frequency $\omega_a$ and a mechanical oscillator (MO) with mass, $m$; frequency, $\omega_m$; and damping rate, $\gamma_m$. This mechanical oscillator is coupled to the cavity mode through the radiation pressure and is simultaneously subjected to an external force $F_{ext}$. Furthermore, this hybrid optomechanical system contains an ensemble of $N$ number of two-level ultracold atoms trapped inside it and interacts non-resonantly with the intracavity field and a classical control field. For a sufficiently large value of $N$, this trapped atomic ensemble behaves effectively as a negative-mass oscillator (NMO) [92, 93, 96–99]. The cavity mode is also coherently driven by a classical field of frequency, $\omega_L$; input power, $P_L$; and wavelength, $\lambda_L$. Moreover, the cavity mode is also coupled to a degenerate OPA. The total Hamiltonian of this hybrid system can be written as the following:

$$\hat{H} = \hat{H}_{OMS} + \hat{H}_{OPA} + \hat{H}_{at} + \hat{H}_L + \hat{H}_F, \quad (1)$$

where $\hat{H}_{OMS}$ describes the cavity optomechanical system, $\hat{H}_{OPA}$ represents the degenerate OPA, $\hat{H}_{at}$ stands for the trapped atomic ensemble, $\hat{H}_L$ accounts for the driving field, and $\hat{H}_F$ denotes the contribution due to the external force. By using the rotating wave approximation and the Holstein–Primakoff transformation for the bosonization approximation for trapped atomic ensemble [92, 93, 96], the total Hamiltonian of this hybrid system in the frame rotating at the driving laser frequency $\omega_L$ can be written as follows:

$$\hat{H}_I = \hbar\Delta_a \hat{a}^\dagger \hat{a} + \frac{\hbar\omega_m}{2}(\hat{X}^2 + \hat{P}^2) + \hbar g_0 \hat{a}^\dagger \hat{a}\hat{X} + i\hbar G(e^{i\theta}\hat{a}^{\dagger 2} - e^{-i\theta}\hat{a}^2) \\ -\hbar\omega_m \hat{d}^\dagger \hat{d} + \frac{\hbar G'}{2}(\hat{a} + \hat{a}^\dagger)(\hat{d} + \hat{d}^\dagger) + i\hbar E_L(\hat{a}^\dagger - \hat{a}) + F_{ext}\hat{X}, \quad (2)$$

where $\hat{a}$ and $\hat{a}^\dagger$ are the annihilation and creation operators of the optical field and $\Delta_a = \omega_a - \omega_L$ is the detuning of the optical mode from the driving laser frequency. Here, $\hat{X}$ and $\hat{P}$ are the dimensionless position and momentum operators of the mechanical oscillator which are normalized to zero-point motion $X_{zp} = \sqrt{\hbar/(m\omega_m)}$ and zero-point momentum $P_{zp} = \hbar/X_{zp}$, respectively. The parameter $g_0$ is the single photon optomechanical coupling strength, whereas $G$ and $\theta$ are the non-linear gain and phase of the degenerate OPA coupled to the cavity mode. Furthermore, $E_L = \sqrt{\kappa P_L/\hbar\omega_L}$ stands for the amplitude of the external driving laser with $\kappa$ as the cavity decay rate. We have used $\hat{d}$ and $\hat{d}^\dagger$ as the effective atomic annihilation and creation operators for the trapped ensemble of the ultracold atoms, whereas $G'$ denotes the collective atomic coupling with the optical cavity mode. The first three terms of Eq. 2 represent the free Hamiltonians of the cavity field, the MO, and the optomechanical interaction between the cavity and the MO. The fourth term represents the coupling between the OPA and the cavity field. The fifth term is the free energy of the trapped atomic ensemble, whereas the sixth term represents the collective interaction of this atomic ensemble with the cavity mode. The seventh term represents the driving term of the cavity mode, and finally, the last term stands for the contribution due to the external force $F_{ext}$ which is to be measured as shown in Figure 1. This external force is also normalized to $\sqrt{\hbar m\omega_m\gamma_m}$ and can be given in the units of $\sqrt{Hz}$. It is worth noting that the atomic ensemble behavior is equivalent to a negative-mass oscillator under the bosonization process, and its frequency is the same as that of the MO, which is the condition of CQNC [79, 80].

## 3 Dynamics of the system

The linearized quantum Langevin equations (QLEs) of this system can be obtained from the Hamiltonian given in Eq. 2 and simultaneously adding the noise and damping terms in these QLEs [64],

$$\dot{\hat{X}} = \omega_m \hat{P}, \\ \dot{\hat{P}} = -\omega_m \hat{X} - g_0 \hat{a}^\dagger \hat{a} - \gamma_m \hat{P} + \sqrt{\gamma_m}(\hat{F}_{th} + F_{ext}), \\ \dot{\hat{a}} = -\left(i\Delta_a + \frac{\kappa}{2}\right)\hat{a} - ig_0 \hat{X}\hat{a} + 2Ge^{i\theta}\hat{a}^\dagger - \frac{iG'}{2}(\hat{d} + \hat{d}') + E_L + \sqrt{\kappa}\hat{a}_{in}, \\ \dot{\hat{d}} = i\omega_m \hat{d} - \frac{iG'}{2}(\hat{a} + \hat{a}^\dagger) - \frac{\Gamma}{2}\hat{d} + \sqrt{\Gamma}\hat{d}_{in}, \quad (3)$$

where $\gamma_m$ and $\Gamma$ are the mechanical damping rate and the collective atomic dephasing rate, respectively. The mechanical noise term is $\hat{F}_{th}$. In the limiting regime of strongly driven cavity field and the atomic ensemble with a weak optomechanical coupling strength, we can linearize these non-linear QLEs given in Eq. 3. So, we expand each operator in Eq. 3 as their mean values plus small quantum fluctuations, i.e., $(\hat{X} = X_s + \delta\hat{X}, \hat{P} = P_s + \delta\hat{P}, \hat{a} = \alpha_s + \delta\hat{a}, \hat{d} = d_s + \delta\hat{d})$. The steady-state values can be obtained by setting all the time derivatives equal to zero in Eq. 3 and are given by

$$X_s = \frac{-g_0 \alpha^2}{\omega_m}, P_s = 0, d_s = \frac{iG'\alpha}{i\omega_m - \frac{\Gamma}{2}}, \\ \alpha_s = \frac{E_L}{\left[\left(i\Delta + \frac{\kappa}{2}\right) - 2Ge^{i\theta} + \frac{iG'^2\omega_m}{\omega_m^2 + \frac{\Gamma^2}{4}}\right]}. \quad (4)$$

Here, $\Delta = \Delta_a + g_0 X_s$ is the effective cavity detuning, and $\alpha_s$ is the steady-state value of the intracavity field amplitude which can always be taken as real and positive with an appropriate chosen phase of the driving field $E_L$. Furthermore, we define the optical and atomic quadrature operators $\hat{x}_a = (\hat{a}^\dagger + \hat{a})/\sqrt{2}$, $\hat{p}_a = i(\hat{a}^\dagger - \hat{a})/\sqrt{2}$, $\hat{x}_d = (\hat{d}^\dagger + \hat{d})/\sqrt{2}$, $\hat{p}_d = i(\hat{d}^\dagger - \hat{d})/\sqrt{2}$, and their corresponding noise operators as $\hat{x}_a^{in} = (\hat{a}^{in,\dagger} + \hat{a}^{in})/\sqrt{2}$, $\hat{p}_a^{in} = i(\hat{a}^{in,\dagger} - \hat{a}^{in})/\sqrt{2}$, $\hat{x}_d^{in} = (\hat{d}^{in,\dagger} + \hat{d}^{in})/\sqrt{2}$, and $\hat{p}_d^{in} = i(\hat{d}^{in,\dagger} - \hat{d}^{in})/\sqrt{2}$ [64]. After these straightforward steps, the linearized quantum Langevin equations for the quadratures fluctuations are obtained in compact form as follows:

$$\delta\dot{\hat{u}}(t) = A\delta\hat{u}(t) + \hat{n}^{in}(t). \quad (5)$$

Here, $\delta\hat{u}^T = (\delta\hat{X}, \delta\hat{P}, \delta\hat{x}_a, \delta\hat{p}_a, \delta\hat{x}_d, \delta\hat{p}_d)$ is the vector of variable operators, $\hat{n}^{in,T}(t) = (0, \sqrt{\gamma_m}(\hat{F}_{th} + F_{ext}), \sqrt{\kappa}\hat{x}_a^{in}, \sqrt{\kappa}\hat{p}_a^{in}, \sqrt{\Gamma}\hat{x}_d^{in}, \sqrt{\Gamma}\hat{p}_d^{in})$ is the vector of noise, and the coefficient matrix $A$ is given as follows:

$$A = \begin{pmatrix} 0 & \omega_m & 0 & 0 & 0 & 0 \\ -\omega_m & -\gamma_m & -g & 0 & 0 & 0 \\ 0 & 0 & c_- & s_+ & 0 & 0 \\ -g & 0 & s_- & -c_+ & -G' & 0 \\ 0 & 0 & 0 & 0 & -\frac{\Gamma}{2} & -\omega_m \\ 0 & 0 & -G' & 0 & \omega_m & -\frac{\Gamma}{2} \end{pmatrix}, \quad (6)$$

where $g = \sqrt{2}g_0\alpha_s$ is effective linear optomechanical coupling strength and its square, i.e., $g^2$ is proportional to the input laser driving power $P_L$, whereas $c_\pm = \pm\frac{\kappa}{2} + 2G\cos\theta$ and $s_\pm = \pm\Delta + 2G\sin\theta$. It can be seen from Eq. 5 that the momentum quadrature $\delta\hat{p}_a$ of the cavity field depends on the position $\delta\hat{X}$ of the mechanical oscillator, and in turn, the position $\delta\hat{X}$ is controlled by the momentum $\delta\hat{P}$ of the mechanical oscillator. Furthermore, the equations of motion given in Eq. 5 can be solved for the operators in the frequency domain to calculate $\delta\hat{p}_a(\omega)$ as a function of the input noises. A quantum operator in the Fourier domain can be written as follows:

$$\delta\hat{O}(\omega) = \frac{1}{\sqrt{2\pi}} \int dt\, \delta\hat{O}(t) e^{-i\omega t}. \tag{7}$$

Therefore, Eq. 5 can be given in the Fourier space as follows:

$$\begin{aligned}
\delta\hat{X} &= \chi_m\left(-g\delta\hat{x}_a + \sqrt{\gamma_m}\left(\hat{F}_{th} + F_{ext}\right)\right), \\
\delta\hat{P} &= \frac{i\omega}{\omega_m}\delta\hat{X}, \\
\delta\hat{x}_a &= \chi'_a\left(s_+\delta\hat{p}_a + \sqrt{\kappa}\,\hat{x}_a^{in}\right), \\
\delta\hat{p}_a &= \chi''_a\Big((g^2\chi_m + s_-)\delta\hat{x}_a - G'\delta\hat{x}_d \\
&\quad - g\chi_m\sqrt{2\gamma_m}\left(\hat{F}_{th} + F_{ext}\right) + \sqrt{\kappa}\,\hat{p}_a^{in}\Big), \\
\delta\hat{x}_d &= \chi_d\left(-\omega_m\delta\hat{p}_d + \sqrt{\Gamma}\,\hat{x}_d^{in}\right), \\
\delta\hat{p}_d &= \chi_d\left(-G'\delta\hat{x}_a + \omega_m\delta\hat{x}_d + \sqrt{\Gamma}\,\hat{p}_d^{in}\right),
\end{aligned} \tag{8}$$

where $\chi'_a$ and $\chi''_a$ are the susceptibilities defined for the cavity field, whereas $\chi_d$ and $\chi_m$ are the susceptibilities of the atomic ensemble and the MO, respectively. Their specific expressions are as follows:

$$\chi'_a = \frac{1}{i\omega - c_-},\ \chi''_a = \frac{1}{i\omega + c_+}, \\ \chi_d = \frac{1}{i\omega + \Gamma/2},\ \chi_m = \frac{\omega_m}{\omega_m^2 - \omega^2 + i\omega\gamma_m}. \tag{9}$$

## 4 Force sensing and CQNC

Solving Eq. 8, we can find the phase quadrature of the cavity field, $\delta\hat{p}_a$, and then using the standard input–output relation as [85].

$$\hat{p}_a^{out} = \sqrt{\kappa}\,\delta\hat{p}_a - \hat{p}_a^{in}. \tag{10}$$

Furthermore, the detected output phase quadrature of the cavity field mode can be expressed in terms of the input noises as follows:

$$\begin{aligned}
\hat{p}_a^{out} &= -g\chi'''_a\chi_m\sqrt{\gamma_m\kappa}\left(\hat{F}_{th} + F_{ext}\right) \\
&\quad + \chi'''_a\left(g^2\chi_m + s_- + G'^2\chi''_d\right)\kappa\chi'_a\hat{x}_a^{in} \\
&\quad + G'\chi'''_a\chi''_d\sqrt{\kappa\Gamma}\left[-\hat{p}_d^{in} + \frac{1}{\omega_m\chi_d}\hat{x}_d^{in}\right] + \left(\chi'''_a\kappa - 1\right)\hat{p}_a^{in},
\end{aligned} \tag{11}$$

where $\chi''_d = -\chi'_d\chi_d^2\omega_m$ and $\chi'_d = (1 + \omega^2\chi_d^2)^{-1}$. The optical susceptibility $\chi'''_a$ can be given as follows:

$$\frac{1}{\chi'''_a} = \frac{1}{\chi''_a} - \left[g^2\chi_m + s_- + G'^2\chi''_d\right]\chi'_a s_+. \tag{12}$$

For applying the CQNC conditions we have considered here

$$g^2\chi_m + G'^2\chi''_d = 0, \tag{13}$$

for all frequencies, and the back-action term will be canceled in Eq. 11, but the small term will remain and contribute as the shot noise which is the effect of the OPA. In the CQNC scheme, we should have $g = G'$ and $\chi_m = -\chi''_d$ or in other words, the contributions to the back-action noise from the mechanical oscillator and the atomic ensemble should cancel each other for all the frequencies [91, 92]. They are the noise and antinoise contributions to the signal obtained with assumption of an atomic ensemble as an oscillator with the NMO. In addition, we have also assumed that the mechanical resonator and the atomic ensemble have the same damping rates ($\gamma_m = \Gamma/2$) and the mechanical oscillator has a high-quality factor such that $\Gamma \ll \omega_m$. In particular, $\gamma_m = \Gamma/2$ and $\Gamma \ll \omega_m$ are conditions of CQNC which were applied in recently published works such as Refs. [76, 77, 91–93]. Therefore, in our work, the susceptibilities due to the mechanical motion and the generalized atomic ensemble will completely match and lead to coherent back-action noise cancellation. By rewriting Eq. 11, the relation between the external force $F_{ext}$ and the measured phase quadrature $\hat{p}_a^{out}$ can be given as follows:

$$F_{ext} + \hat{F}_{add} = \frac{\hat{p}_a^{out}}{-g\chi'''_a\chi_m\sqrt{\gamma_m\kappa}}, \tag{14}$$

whereas the added force noise $\hat{F}_{add}$ is given by

$$\begin{aligned}
\hat{F}_{add} &= \hat{F}_{th} - \frac{\chi'''_a\kappa - 1}{g\chi'''_a\chi_m\sqrt{\gamma_m\kappa}}\hat{p}_a^{in} - \frac{\kappa\chi'_a s_-}{g\chi_m\sqrt{\gamma_m\kappa}}\hat{x}_a^{in} \\
&\quad + \left[-\hat{p}_d^{in} + \frac{i\omega + \Gamma/2}{\omega_m}\hat{x}_d^{in}\right].
\end{aligned} \tag{15}$$

The aforementioned Eq. 15 contains the thermal Langevin force coupled with a thermal reservoir at temperature $T$ (first term), the shot noise in the phase quadrature of the optical field (second term), the noise injected into the system through the OPA (third term), and the atomic noise (fourth term). Furthermore, we can find the sensitivity of the force measurement by using the definition of spectral density $S_{F,add}(\omega)$ of the added noise [77, 100] as follows:

$$S_{F,add}(\omega)\delta(\omega - \omega') = \frac{1}{2}\left(\langle\hat{F}_{add}(\omega)\hat{F}_{add}(-\omega')\rangle + c.c.\right). \tag{16}$$

Using the conditions of the CQNC and for $\omega \ll \kappa$ (this is not difficult to achieve experimentally, since it is easy to prepare bad cavities), this approximation is also used in earlier works [91–93, 96]; further, we find the spectral noise as follows:

$$\begin{aligned}
S_{F,add}(\omega) &= \frac{k_B T}{\hbar\omega_m} + \frac{1}{g^2|\chi_m|^2(\gamma_m\kappa)}\left[\left|c_+ + \frac{s_+ s_-}{c_-}\right|^2 + \kappa^2\left|\frac{s_-}{c_-}\right|^2\right] \\
&\quad + \frac{1}{2}\frac{\omega^2 + \omega_m^2 + \Gamma^2/4}{\omega_m^2},
\end{aligned} \tag{17}$$

where the back-action noise term scaled to $g^2$ is notably deleted. Furthermore, our obtained result can be compared to the standard optomechanical system, discussed in Refs. [1, 77, 93], as the following equation

$$S_F(\omega) = \frac{k_B T}{\hbar\omega_m} + \frac{1}{2}\left[\frac{\kappa}{\gamma_m}\frac{1}{g^2|\chi_m|^2}\frac{1}{4} + 4g^2\frac{1}{\kappa\gamma_m}\right], \tag{18}$$

which contains the shot noise and the back-action noise terms proportional to $1/g^2$ and $g^2$, respectively. The thermal Brownian noise results in a constant background noise to the force sensitivity independent of the input laser power. At $T = 0$, we can minimize the right hand side of Eq. 18 with respect to $g^2$, or proportionally, the input laser power $P_L = 2\hbar\omega_L\kappa(g/g_0)^2$, gives the SQL, the achievable lower bound, for weak force sensing,

$$S_{F,SQL} = \frac{1}{\gamma_m|\chi_m|}, \tag{19}$$

and equivalently, the minimized noise spectral density for the CQNC at $T = 0$ is given as follows:

$$S_{F,CQNC} = \frac{1}{2}\frac{\omega^2 + \omega_m^2 + \Gamma^2/4}{\omega_m^2}. \tag{20}$$

We will compare the noise spectrum in our proposed scheme with a bare standard optomechanical system formed by an optical cavity coupled with a MO [1]. The SQL for stationary force detection comes from the minimization of the force spectrum of the standard optomechanical system at a given frequency over the driving power as given in Refs. [1, 92, 93, 96]. Next, we use a set of parameters to simulate the noise spectral densities of the current research system, the fixed parameters are given in Table 1, and the rest of the parameters are taken as the phase of the OPA, $0 \leqslant \theta \leqslant 2\pi$; the gain of OPA, $0 \leqslant G \leqslant 0.3\kappa$; and also three values for detuning, $\Delta/\omega_m = 0, 1, 2$.

TABLE 1 Parameters used in our proposed hybrid optomechanical system [64, 76, 77, 93, 101].

| Parameter | Symbol | Value (unit) |
|---|---|---|
| Mirror mass | $m$ | 50 ng |
| Single-photon optomechanical coupling | $g_0/2\pi$ | 300 Hz |
| Mechanical resonance | $\omega_m/2\pi$ | 300 kHz |
| Mechanical damping rate | $\gamma_m/2\pi$ | 30 Hz |
| Optical cavity damping rate | $\kappa/2\pi$ | 1 MHz |
| Laser source power | $P_L$ | 100 mW |
| Laser frequency | $\omega_L/2\pi$ | 384 THz |

The noise spectral densities for the standard optomechanical system (SQL, black solid line), the proposed hybrid optomechanical system with OPA, and atomic ensemble (dashed lines) including CQNC schemes (red solid line) are shown in Figure 2, where the black solid line is obtained using Eq. 18, the dashed lines, namely, $G/\kappa = 0, 0.1, 0.3$, and $\theta = 0, \pm\pi/2, \pm\pi, \pm3\pi/2$ correspond to Eq. 17, and the solid red line corresponds to Eq. 20. The spectral density for the CQNC scheme is limited to the shot noise term over the whole detection bandwidth and is the main advantage of the CQNC scheme. In this context, the variation of the noise spectral density of the hybrid optomechanical system with the detection frequency for the different values of the OPA pump gain $G$ and the phase angle $\theta = \pi$ is shown in Figure 2A. At the mechanical resonance condition $\left(\omega = \omega_{m}\right)$ and for effective cavity detuning $\Delta = 0$, the noise spectral density of the standard optomechanical system (SQL) and that of the hybrid optomechanical system with the OPA pump is equal to the noise spectral density of the CQNC scheme. With the gradual increase in the OPA pump gain, $G$, the noise spectral density can be suppressed almost nearly by two orders of magnitude at frequencies below and above the mechanical resonance condition as shown in Figure 2A. In addition, we find that in the case of $G = 0$, the dashed purple and the black solid line almost coincide. This is because, in this case, the first two terms in both Eqs 17, 18 are much bigger than their later terms, and consequently, they have almost the same behavior. Then, we have the same result obtained by a conventional CQNC scheme [93]. However, if we take non-zero values for $G$, we have smaller values for $S_F$ as can be seen for $G = 0.1$ and $G = 0.3$ in Figure 2A. In fact, it shows that present scheme has much better results than conventional CQNC scheme such as those in Ref. [93]. Furthermore, an experimental scheme to expand the detection bandwidth for gravitational-wave observation in which quantum uncertainty can be squeezed inside one of the optical resonators at high frequencies while keeping the low-frequency sensitivity unchanged [102]. However, for a given value of $G$, the noise spectral density gets suppressed only for $\theta = \pi$, whereas for other values of $\theta$, its values increase beyond the SQL as shown in Figure 2B.

The noise spectral densities as functions of the laser driving power $P_L$ (scaled to the optomechanical coupling strength $g^2$) for on-resonance $(\omega = \omega_m)$ and off-resonance $(\omega = \omega_m + 4\gamma_m)$ conditions are, respectively, shown in Figure 3. At the higher driving power, for both these cases, the noise spectral density of our hybrid optomechanical system decreases significantly over the bandwidth. However, for lower values of the OPA gain $G$, the shot noise term is still dominant. The noise spectral density of the normal optomechanical system (SQL) first decreases up to a minimum point with the driving power; afterward, the back-action noise is dominant and it again increases. The noise spectral density of the SQL represented by Eq. 18 can reach the minimum point for $g_{SQL} = \sqrt{\kappa}/(2\sqrt{|\chi_m|})$, whereas in this hybrid optomechanical system, it can be lowered by using the OPA pump with gain $G$ and given as $g_{SQL}^{\theta=\pi} = |\kappa - 4G|/(2\sqrt{\kappa|\chi_m|})$, including complete elimination of the shot noise due to the presence of trapped atomic ensemble acting as a negative-mass

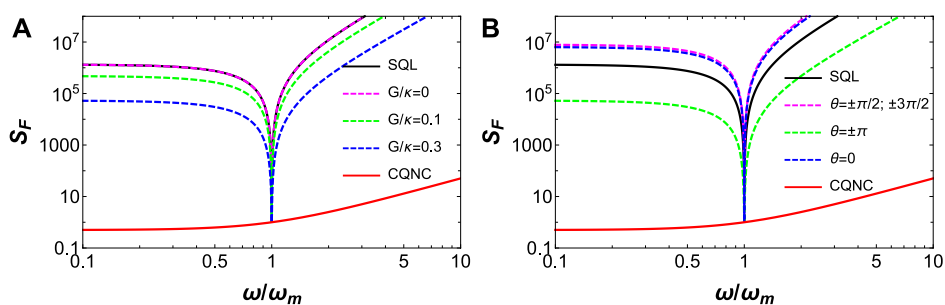

FIGURE 2
Noise power spectral densities for the standard optomechanical system $S_{SQL}$ (black solid line), the hybrid optomechanical system with the OPA (colored dashed line), and the coherent quantum noise cancellation $S_{CQNC}$ (red solid line) as a function of frequency $\omega/\omega_m$. The spectral densities are normalized to $\hbar m \omega_m \gamma_m$. The OPA parameters are given as (A) $\theta = \pi$ and (B) $G/\kappa = 0.3$ and the rest of the parameters are given in Table 1. Here, we have taken $Q = \omega_m/\gamma_m = 10^4$.

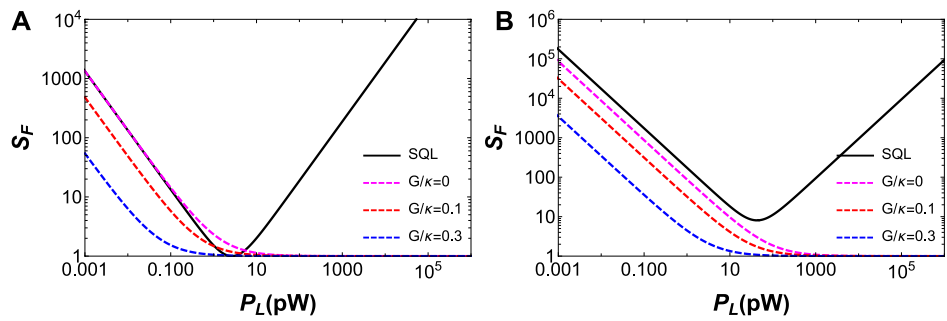

FIGURE 3
Noise spectral density **(A)** on mechanical resonance ($\omega = \omega_m$) and **(B)** off-resonance ($\omega = \omega_m + 4\gamma_m$) as a function of the input driving power $P_L$ for the standard optomechanical system (SQL) and for the present hybrid optomechanical system with different values of OPA gain $G$ and for $\Delta = 0$, $\theta = \pi$. The other parameters are given in Table 1.

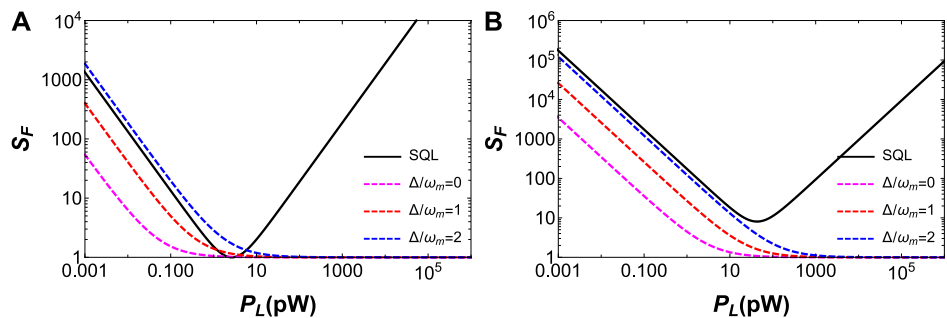

FIGURE 4
Noise spectral density **(A)** on mechanical resonance ($\omega = \omega_m$) and **(B)** off-resonance ($\omega = \omega_m + 4\gamma_m$) as a function of the input driving power $P_L$ for the standard optomechanical system (SQL) and for the present hybrid optomechanical system with different values of the effective cavity detuning $\Delta$ and for $G/\kappa = 0.3$, $\theta = \pi$. The other parameters are given in Table 1.

oscillator. In the absence of OPA pumping ($G = 0$), for both on-resonance and off-resonance cases, this hybrid optomechanical system needs a higher driving power to reach the same sensitivity given by the SQL. However, with the gradual increase in the OPA pump gain $G$, the sensitivity of this CQNC–OPA arrangement becomes superior to the normal optomechanical system and surpasses the SQL at lower values of the driving power $P$, thus also improving the measurement accuracy at the lower driving power domain. Furthermore, this proposed OPA-based CQNC scheme cancels the back-action noise completely and also reduces the shot noise by almost two orders of magnitude as shown in Figures 3A, B. In addition, this hybrid optomechanial system also improves the sensitivity in a broader bandwidth range even at lower values of the driving power for a given set of physical parameters already used in Refs. [64, 77, 93] and given in Table 1.

In Figure 4, we have plotted the noise spectral density of force measurement in mechanical on-resonance and off-resonance conditions for different values of the effective detuning, $\Delta$. For on-resonance case ($\omega = \omega_m$), it can be seen that resonant CQNC with $\Delta = 0$ gives better sensitivity for force measurement than heterodyne CQNC with $\Delta = \omega_m$ and $\Delta = 2\omega_m$, as shown in Figure 4A. Moreover, for a higher value of cavity detuning, $\Delta = 2\omega_m$, noise spectral density can even surpass the SQL, and so, it degrades force sensitivity. Furthermore, for off-resonance case ($\omega = \omega_m + 4\gamma_m$), again, $\Delta = 0$ gives better force sensitivity as compared to heterodyne CQNC as shown in Figure 4B. However, this resonant CQNC gives much better results for on-resonance cases only. This is because in case of on-resonance condition, the anti-Stokes process dominates within the system which leads to a significant cooling of the mechanical oscillator and facilitates a better force sensitivity.

We have studied the effect of $g^2/g_0^2$ on the noise spectral density of the present hybrid system, as shown in Figure 5. It can be seen that in absence of the OPA, i.e., $G = 0$ as $g^2/g_0^2$ (which is proportional to the input laser power $P_L$) increases; we get the minimum value of noise spectral density only in the lower detection frequency range $\omega$ which is very far away from mechanical resonance condition, and after that, its value increases. It means that in the absence of the OPA, better measurement accuracy of the external weak force is obtained only at a lower detection frequency range. When $G$ is not 0, that is, in the presence of the OPA, comparing the values of the color area and color bar in Figures 5A, B, we find that the OPA causes the noise spectral density to decrease in the entire parameter area. In this case, even smaller values of $g^2/g_0^2$ or, in other words, a smaller input laser power $P_L$, can lead to these values of noise spectral density, which means the presence of the OPA significantly reduces the noise spectral density in the entire parameter area and hence, significantly improves the force sensitivity. Figure 5C shows that, in case of a fixed value of the OPA gain $G$, the noise spectral density is symmetric about $\theta = 0$, which indicates that the positive and negative values of $\theta$ do not affect the measurement accuracy and only depends upon the absolute magnitude of $\theta$. We have also found that when $|\theta|$ is large, much smaller values of $g^2/g_0^2$ can achieve higher sensitivity of weak force, whereas for $\theta = 0$, the cost is to increase $g^2/g_0^2$ to achieve better measurement accuracy.

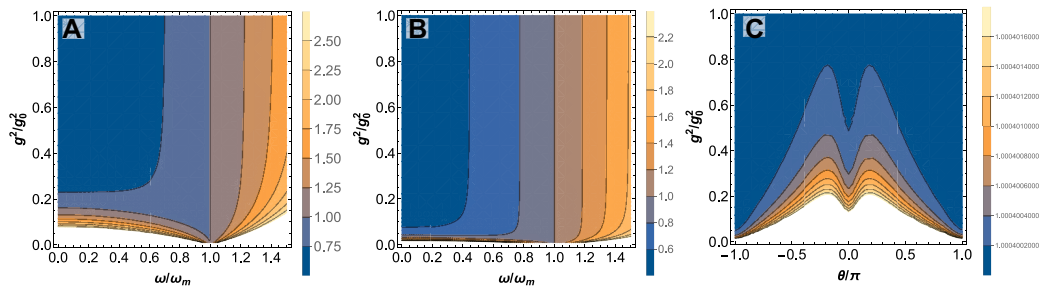

FIGURE 5
Contour plot of noise spectral density for weak force measurement as function of **(A)** the relative squared optomechanical coupling strength ($g^2/g_0^2$) and the relative detection frequency ($\omega/\omega_m$) for $G/\kappa = 0$, **(B)** for $G/\kappa = 0.2$ and $\theta = \pi$ **(C)** with the phase of the OPA $\theta$ for $G/\kappa = 0.2$ and $\omega = \omega_m$. The other parameters are given in Table 1.

Based on our research results, we would like to add here that unlike earlier work cited in Ref. [96] where all the numerical results were studied for only one optimal value of squeezing phase minimizing the shot noise contribution, we have explored the most generalized scenario with different non-linear gain and phase angle of the OPA; in fact, in our scheme, both the parameters of OPA can significantly reduce the measurement of weak force sensing. So, our numerical results are more generalized and are easily implemented in comparison to that of Ref. [96]. This is because, as in the work given by Ref. [96], they need an infinite bandwidth value and very high squeezing parameters for the externally coupled squeeze vacuum reservoir to the cavity mode; this is still very challenging from the experimental and integrated photonics points of view as their scheme cannot avoid the detrimental effects of the inevitable major losses in the transmission and injection of the squeezed light from the external squeezed reservoir into the optomechanical cavity. Furthermore, from the analytical point of view, these squeezing bandwidth parameters related to the external vacuum reservoir should ideally be infinite; only then will the Markovian approximation hold valid for the input noise terms related to cavity field operators in Ref. [96], whereas in our case, OPA parameters are almost of the order of the cavity decay rate which again makes our scheme easily feasible in future experiments. Another important advantage of adding squeezed OPA within an optomechanical system is that it provides significant ground state cooling as shown in our earlier work [24], which is very difficult for the scheme given in Ref. [96], and hence, our scheme will give better quantum measurement due to the cooling effect even in unresolved sideband regime of cavity optomechanics.

## 4.1 Experimental feasibility and measurement of weak force

Finally, it is necessary to discuss the experimental feasibility and detection of the weak force in the cavity optomechanical system. Here, we need to admit that implementing our proposal experimentally is challenging. Fortunately, Hertzberg et al. demonstrated back-action-evading measurements of a single quadrature of nanomechanical motion can increase the force sensitivity [74]. Furthermore, Møller et al. in 2017 demonstrated destructive or constructive interference of the quantum back-action for the two mechano-oscillators, further showing that the back-action-evading measurement in the hybrid system leads to an enhancement of displacement sensitivity in a negative mass reference frame [103]. This suggests that the quantum back-action noise can be eliminated experimentally by some quantum physical mechanism. In addition, a recent study reported that they observed quantum back-action noise due to optomechanical coupling leading to correlated mechanical fluctuations of the two mechanical oscillators in a driven optical cavity [104]. The observed quantum back-action noise undoubtedly provides a solid foundation for the realization of our scheme. Moreover, Daniel et al. reported that the collective motion of atoms can be driven through quantum noises in the radiation pressure and that the quantum back-action of this motion onto the cavity field produces ponderomotive squeezing. In turn, they detected this quantum phenomenon by measuring sub-wave-noise optical squeezing [105].

In particular, the ensembles of ultracold atoms are used for atomic clocks [106], quantum simulation [107], quantum information processing [108], dynamical phase transitions in an optical cavity [109], and coherent quantum noise cancellation [93, 96]. According to the purpose in these works, a large range of the number of atoms has been used. In the optomechanical CQNC schemes, in one hand the cavity field is strongly driven but on the other hand there is a weak coupling between the atom-field modes which consequently there is the low-excitation limit for atomic ensemble. In fact, under the conditions of large atom number N, weak atom-cavity coupling, and the low-excitation limit, the dynamics of the atomic ensemble can be described in terms of a collective bosonic operators. For example, the total number of atoms are approximately large such as $N \approx 1 \times 10^{13}$ used to show Einstein-Podolsky-Rosen paradox [110], and $N \approx 1 \times 10^{13}$ used in the CQNC force sensing [93] as we have assumed in the present scheme. Note that the number of excited atoms can be quantified by $\langle d^\dagger d \rangle = |d_s|^2$. Under our current parameters ($P_L = 10^{-3}$pW ~ $10^6$pW), the number of excitations of the atoms $\langle d^\dagger d \rangle = 10^5$. A standard number of atomic ensemble is $N = 1 \times 10^8$, so $\langle d^\dagger d \rangle / N = 10^8 = 10^{-3} \ll 1$, that is the Holstein-Primakoff mapping approximation is completely valid.

On the other hand, quantum systems that exhibit dynamic quantum back-action of radiation pressure have to address a range of design considerations, including physical size and dissipation [2]. Fortunately, recent experimental progress in the fabrication technology of cavity optomechanical systems is very mature and perfect, opening up enormous possibilities in the design and integration of hybrid quantum interfaces based on optical cavities [111], for example, the optomechanical resonator with a micromechanical membrane coupled simultaneously with atomic gas [112, 113]. In particular, this also lays the foundation for the conditions under which CQNC can be realized. In addition, these hybrid systems that combine optomechanical resonators with the best features of atomic (or atom like) systems to develop experimentally feasible approaches based on CQNC are given in Refs. [89, 103].

In order to better promote the experiment realization of the present scheme, we briefly give the reference ranges for the key parameters in the experiment according to the current experimental ability. It is worth noting that these parameter values are mainly collected by Aspelmeyer et al. [1]. When selecting experimental parameters, we need to comprehensively consider the stability of the system, the difficulty of device processing, the experimental cost, etc. Specifically, the mass of the moving mirror $m$ is chosen as $10^{-22}$ kg ~ $1.9 \times 10^{-7}$ kg [114–116], the single photon optomechanical coupling ($g_0/2\pi$) is 1.2 Hz ~ $9 \times 10^5$ Hz [115, 117], the mechanical frequency ($\omega_m/2\pi$) is $9.7 \times 10^3$ Hz ~ $3.9 \times 10^9$ Hz [117, 118], the mechanical damping rate ($\gamma_m/2\pi$) is $1.3 \times 10^{-2}$ Hz ~ $3.9 \times 10^4$ Hz [117, 118], and the optical cavity damping rate ($\kappa/2\pi$) is $2 \times 10^5$ Hz ~ $3.9 \times 10^8$ Hz [117, 119].

## 5 Conclusion

We theoretically investigate the weak force sensing through the CQNC scheme in a hybrid optomechanical system containing an ensemble of trapped ultracold atoms and the OPA. For a sufficiently large number of atoms, this trapped ensemble of ultracold atoms acts as a negative-mass oscillator, which destructively interacts with the optical cavity mode leading to the cancellation of the back-action noise. In addition, the presence of OPA reduces the shot noise in the regime of the low driving power and also gives a broad range of detection frequency. In particular, the CQNC and shot noise reduction occur when the effective linear optomechanical coupling strength $g$ and the collective atomic coupling with the optical cavity mode given by $G$ are both equal to each other. Furthermore, it can be seen that in case of resonance CQNC, i.e., effective cavity detuning $\Delta = 0$, a comparable value of $G$, and suitable value of phase angle $\theta$ leads to the suppression of the noise spectral density significantly at frequencies below and above the mechanical resonance condition. Our study provides a promising platform for weak force sensing and can also be explored in other systems of quantum sensing with waveguide, interferometer, or parity-time symmetric microcavity.

## Data availability statement

The original contributions presented in the study are included in the article/Supplementary Material, further inquiries can be directed to the corresponding author.

## Author contributions

All authors listed made a substantial, direct, and intellectual contribution to the work and approved it for publication.

## Acknowledgments

MK thanks and acknowledges the International Research Network Grant Scheme (STR-IRNGS-SET-GAMRG-01-2021).

## Conflict of interest

The authors declare that the research was conducted in the absence of any commercial or financial relationships that could be construed as a potential conflict of interest.

## Publisher's note

All claims expressed in this article are solely those of the authors and do not necessarily represent those of their affiliated organizations, or those of the publisher, the editors, and the reviewers. Any product that may be evaluated in this article, or claim that may be made by its manufacturer, is not guaranteed or endorsed by the publisher.